# Negative c-axis magnetoresistance in graphite

Y. Kopelevich[1, 2], R. R. da Silva[1], J. C. Medina Pantoja[1], and A. M. Bratkovsky[2]

[1]Instituto de Física "Gleb Wataghin", Universidade Estadual de Campinas, UNICAMP 13083-970, Campinas, São Paulo, Brasil

[2]Hewlett-Packard Laboratories, 1501 Page Mill Road, Palo Alto, California 94304

ABSTRACT

We have studied the c-axis interlayer magnetoresistance (ILMR), $R_c(B)$ in graphite. The measurements have been performed on strongly anisotropic highly oriented pyrolytic graphite (HOPG) samples in magnetic field up to B = 9 T applied both parallel and perpendicular to the sample c-axis in the temperature interval 2 K ≤ T ≤ 300 K. We have observed negative magnetoresistance, $dR_c/dB < 0$, for B ∥ c-axis above a certain field $B_m(T)$ that reaches its minimum value $B_m$ = 5.4 T at T = 150 K. The results can be consistently understood assuming that ILMR is related to a tunneling between zero-energy Landau levels of quasi-two-dimensional Dirac fermions, in a close analogy with the behavior reported for α-(BEDT-TTF)$_2$I$_3$ [N. Tajima et al., Phys. Rev. Lett. **102**, 176403 (2009)], another multilayer Dirac electron system.

Graphite consists of N > 1 layers of carbon atoms packed in honeycomb lattice, dubbed graphenes, with two non-equivalent sites, A and B, in the Bernal (ABAB...) stacking configuration [1]. Conducting (conjugated) π-electrons move within planes, formed by $p_z$-wave functions density maxima located parallel (above and below) to graphene layers. In the absence of interlayer electron hopping, charge-carrying quasi-particles (QP) have the linear dispersion relation $E(p) = \pm v|p|$, and the Fermi surface is reduced to two points (K and K′) at the opposite corners of the 2D hexagonal Brillouin zone. Such carriers can be described as massless (2+1)D Dirac fermions (DF) [2] providing a link to relativistic models for particles with an effective "light" velocity $v \approx 10^6$ ms$^{-1}$. On the other hand, according to Slonczewski-Weiss-McClure (SWMC) model [1], the interlayer coupling leads to a dispersion in the $p_z$-direction with cigar-like electron (K) and hole (H) Fermi surface pockets elongated along the corner edge HKH of a 3D Brillouin zone, and no linear dispersion is expected, except very close to the H point [3]. However, studies of Shubnikov de Haas (SdH) and de Haas van Alphen (dHvA) quantum oscillations [4] showed that DF in graphite occupied an unexpectedly large phase volume. The DFs in graphite were also detected by means of angle-resolved photoemission spectroscopy (ARPES) [5], scanning-tunneling-spectroscopy (STS) [6], and infrared magneto-transmission [7] techniques. Besides, micro-Raman [8], STS [9] and microwave magneto-absorption [10] measurements provided evidence for the existence of independent (decoupled) graphene layers in bulk graphite.

The question whether graphene layers are situated only at the sample surface or they are distributed through the sample volume remains unclear. Aiming to verify the possible multi-graphene behavior of graphite, we have studied in the present work the c-axis interlayer magnetoresistance (ILMR). In particular, it is expected that ILMR is dominated by interlayer tunneling between zero-energy Landau levels of quasi-2D DF, and the perpendicular magnetic field, increasing the zero energy state degeneracy, results in the resistance decreasing with the field, i. e. negative ILMR (NILMR) [11]. Here we report the observation of NILMR in our most anisotropic graphite samples that resembles the resistance behavior reported for α-(BEDT-TTF)$_2$I$_3$ [12], another multilayer Dirac electron system.

The measurements were performed on thoroughly characterized [13] HOPG-UC (Union Carbide Co.) with the room temperature, zero-field, out-of-plane/basal-plane resistivity ratio $\rho_c/\rho_b = 3 \cdot 10^4$, $\rho_c = 0.1$ Ωcm, and mosaicity of 0.4° (FWHM obtained from x-ray rocking curves). X-ray diffraction (Θ-2Θ) spectra revealed a characteristic hexagonal graphite structure in the Bernal (ABAB…) stacking configuration, with no signature of the rhombohedral phase. The obtained crystal lattice parameters are a = 2.48 Å and c = 6.71 Å. Both dc and low-frequency (f = 1 Hz) ac resistance measurements were made using commercial He-4, B = 9 T cryostats. For the measurements, silver paste electrodes were placed on the sample surface(s). The c-axis resistance $R_c(B,T)$ was measured by attaching two electrodes to each of the main (basal) sample surfaces; one is point-like in the middle and other surrounding it and contacted on the rest surface, assuring a uniform current distribution. Complementary measurements of the in-plane resistance $R_b(B, T)$ were performed by attaching four contacts to the same sample surface. Measurements were performed for both B ∥ c and B ⊥ c configurations. The results presented below were obtained for the sample with dimensions l x w x t = 5 x 5 x 1 mm$^3$ (t ∥ c-axis).

Figure 1 (a, b) presents $R_c(B)$ isotherms recorded for various temperatures. The salient feature of the data, i. e. the maxima in $R_c(B)$, is indicated by arrows. The negative magnetoresistance ($dR_c/dB < 0$) reveals itself at $B > B_m(T)$. As shown in Fig. 2, $B_m(T)$ is a non-monotonic function having the minimum at T ~ 150 K.

Taking into account that the basal-plane MR in well graphitized samples is positive and very big [1], see also the inset in Fig. 1(a), the occurrence of negative MR in the certain domain of the B-T plane (Fig. 2) should be specific to the c-axis transport. In order to clarify its origin, we first turn our attention to the zero-field anisotropic electrical transport.

Figure 3 presents normalized $\rho_c(T)$ and $\rho_b(T)$ resistivity curves obtained for B = 0. As can be seen from Fig. 3, $\rho_c(T)$ demonstrates insulating-like ($d\rho_c/dT < 0$) behavior for T > 50 K, whereas $\rho_b(T)$ rapidly decreases for T < 150 K ($d\rho_b/dT > 0$). As a result, the anisotropy $\rho_c/\rho_b$ increases as the temperature decreases (see the inset in Fig. 3), representing the characteristic feature of most anisotropic HOPG. The ratio $\rho_c/\rho_b$ vs. T can be best fitted by the equation:

$$\rho_c/\rho_b = a + bT^{-\alpha}, \qquad (1)$$

where $\alpha = 0.5$. The insulating-type $\rho_c(T)$ is the inherent property of a disorder-free graphite, but the presence of lattice defects can act as short-circuits between the graphene planes, reducing the measured $\rho_c(T)$ compared to its true value [1]. Recent theoretical studies of the perpendicular transport in graphene multi-layers [14] led to similar conclusions: (i) the perpendicular transport is enhanced by structural disorder; (ii) the cleaner the system the larger the anisotropy; (iii) the resistivity ratio $\rho_c/\rho_b \sim T^{-\alpha}$ at the Dirac point, and diverges when T → 0. However, the saturation of $\rho_c/\rho_b$ with the temperature lowering is expected for finite doping (chemical potential), being in agreement with our observations (see the inset in Fig. 3). The predicted exponent $\alpha = 0.66(6)$ is also close to the experimental value $\alpha = 0.5$ found in the present work.

Because of the structural disorder, the current in the measurements flows not only perpendicular but also parallel to graphene planes. This explains the occurrence of SdH oscillations in the B || I || c configuration, clearly seen for T = 2 K and T = 10 K in Fig. 1(a), top curve, as well as the metallic-like behavior of $R_c(T)$ at T < 50 K (Fig. 3): both effects are likely governed by $R_b(T)$.

Then, the resistance measured along the c-axis can be described by the equation for two parallel resistors $R_{eff} = R_c R_b/(R_c+R_b)$, where $R_b(B) = R_0 + aB^{s(T)}$ [1, 15] and

$$R_c(B) = A/(B + B_0), \qquad (2)$$

predicted for multilayer DF systems [11, 12], where A and $B_0$ are fitting parameters. The background physics behind the decreasing $R_c$ vs. B (NILMR) is the Dirac-type Landau level quantization $E_n = \pm (2e\hbar v_F^2|n|B)^{1/2}$ [2], where the lowest Landau level is located precisely at $E_0 = 0$ (zero mode). This has an important consequence: when the interlayer transport is governed by the tunneling between zero-energy Landau levels, the increase of zero-mode degeneracy with the field leads to NILMR [11, 12]. Red lines in Fig. 1 are obtained from the equation for $R_{eff}(B)$ exemplifying good agreement with the experimental results. Somewhat similar interpretation of the field-driven crossover from positive to negative MR has recently been proposed in the context of α-(BEDT-TTF)$_2$I$_3$ [16], assuming a non-vertical interlayer tunneling. According to Ref. [16], the crossover field $B_m$ that marks a maximum in $R_c(B)$ corresponds to the crossover from the inter-Landau level mixing regime (B ≤ $B_m$) to the interlayer tunneling regime between well separated zero-energy Landau levels (B ≥ $B_m$). This implies that for B > $B_m$, one has $E_1 - E_0 > k_B T$, Γ, $t_c$ where Γ = $\hbar/\tau$ (τ is the transport relaxation time) and $t_c$ measure the strength of a quenched disorder and interlayer tunneling, respectively. It is worth noting, that the old SWMC model for graphite suggests $t_c \approx 0.39$ eV $\gg k_B T$, and at first glance the comparison of our experimental results with the theoretical models [11, 16], where $t_c < k_B T$, may not be appropriate. However, this value of $t_c$ is nearly two orders of magnitude larger than the value ~ 5 meV reported e. g. by Haering and Wallace [17] who pointed out the 2D character of QPs in graphite, see also Refs. [4, 18]. Also, recent measurements [19] of current-voltage (I-V) characteristics performed on graphitic mesas suggested the interlayer tunneling of Dirac fermions between Landau levels. It seems, both energy scales are relevant in graphite which can be viewed as the stack of alternating "blocks" of strongly and weakly coupled graphene planes [18, 20], so that the NILMR originates from the tunneling between nearly decoupled graphitic planes. Taking characteristic $\tau(T)$ for graphite [21], the inequality $(E_1 - E_0)/k_B[K] \approx \pm 420|n|^{1/2}(B[T])^{1/2} > \{T, \Gamma\}$ is satisfied over the whole NILMR domain on the B-T plane in Fig. 2.

From this perspective, one also understands the non-monotonic $B_m(T)$. As the temperature decreases from 300 K to ~ 100 K, the condition $E_1 - E_0 > \{k_B T, \Gamma\}$ improves and hence $B_m$ decreases. At the same time, the inequality $t_c < k_B T$ inevitably inverts as $T \to 0$, implying the divergence of $B_m$ with the temperature lowering as Fig. 2 illustrates. Assuming that the minimum in $B_m(T)$ corresponds to the condition $t_c \approx k_B T$, one gets $t_c$ ~ 10 meV, close to the value obtained in Ref. [17].

Testing further the theoretical model for NILMR [11], we performed $R_c(B)$ measurements with B ∥ basal planes (Fig. 4). As Fig. 4 illustrates, in this geometry the negative MR does not occur for all studied temperatures and magnetic fields, providing evidence that NILMR is indeed associated with the field component perpendicular to basal planes.

Summarizing, we report the first observation of negative interlayer magnetoresistance (NILMR) in strongly anisotropic highly ordered graphite that can be consistently understood within a framework of tunneling models between zero-energy Landau levels of quasi-two-dimensional Dirac fermions [11, 16]. This finding together with the zero-field resistivity anisotropy measurements, performed in this work, provides an additional experimental evidence that graphite consists (at least partially ) of weakly coupled graphene planes with the Dirac-type quasiparticle spectrum.

This work was partially supported by FAPESP, CNPq, and INCT NAMITEC.

[16] T. Morinari and T. Tohyama, arXiv:0912.0566 [cond-mat. mtrl-sci].

[17] R. R. Hearing and P. R. Wallace, J. Phys. Chem. Solids **3**, 253 (1957).

[18] Y. Kopelevich, P. Esquinazi, J. H. S. Torres, R. R. da Silva, and H. Kempa, Adv. Solid State Phys. **43**, 207 (2003); Y. Kopelevich and P. Esquinazi, Adv. Mater. **19**, 4559 (2007).

[19] Yu. I. Latyshev, Z. Ya. Kosakovskaya, A. P. Orlov, A.Yu. Latyshev, V.V. Kolesov, P. Monceau, and D. Vignolles, Journal of Physics: Conference Series **129 ,** 012032 (2008) .

[20] I. A. Luk'yanchuk and Y. Kopelevich, Phys. Rev. Lett. **97**, 256801 (2006).

[21] L. C. Olsen, Phys. Rev. B **6**, 4836 (1972).
9

**FIGURE CAPTIONS**

Fig.1. (Color online) $R_c(B)$ isotherms recorded for various temperatures. The negative magnetoresistance ($dR_c/dB < 0$) takes place for $B > B_m(T)$ indicated by arrow. Red lines are obtained from the equation $R_{eff} = R_c R_b/(R_c+R_b)$, where $R_b(B) = R_0 + aB^{s(T)}$ is the basal-plane resistance contribution (see text) and $R_c(B) = A/(B + B_0)$ with $R_0$, a, A, and $B_0$ being fitting parameters: $R_0 = 0.15$ $\Omega$, a = 0.26 $\Omega \cdot T^{-1.27}$, A = 19 $\Omega \cdot T$, $B_0 = 10$ T, s = 1.3 (T = 50 K); $R_0 = 0.085$ $\Omega$, a = 0.1 $\Omega \cdot T^{-1.5}$, A = 4.85 $\Omega \cdot T$, $B_0 = 10$ T, s = 1.5 (T = 100 K); $R_0 = 0.08$ $\Omega$, a = 0.04 $\Omega \cdot T^{-1.6}$, A = 2.5 $\Omega \cdot T$, $B_0 = 11$ T, s = 1.6 (T = 175 K). The inset in (a) exemplifies $R_b(B)/R_b(0)$ measured with the current flowing parallel to basal graphitic planes.

Fig.2. $B_m(T)$ separates positive ($dR_c/dB > 0$) and negative ($dR_c/dB < 0$) magnetoresistance.

Fig.3. (Color online) Normalized c-axis $\rho_c(T)/\rho_c(300$ K$)$ and basal-plane $\rho_b(T)/\rho_b(300$ K$)$ resistivity curves obtained for B = 0. The inset presents the anisotropy $\rho_c/\rho_b$ vs. T; the red line is obtained from Eq. (1): a = 1.1, b = 40 $K^{\alpha}$, $\alpha = 0.5$.

Fig.4. $R_c(B)$ measured at various temperatures and B || basal planes.

Fig. 1

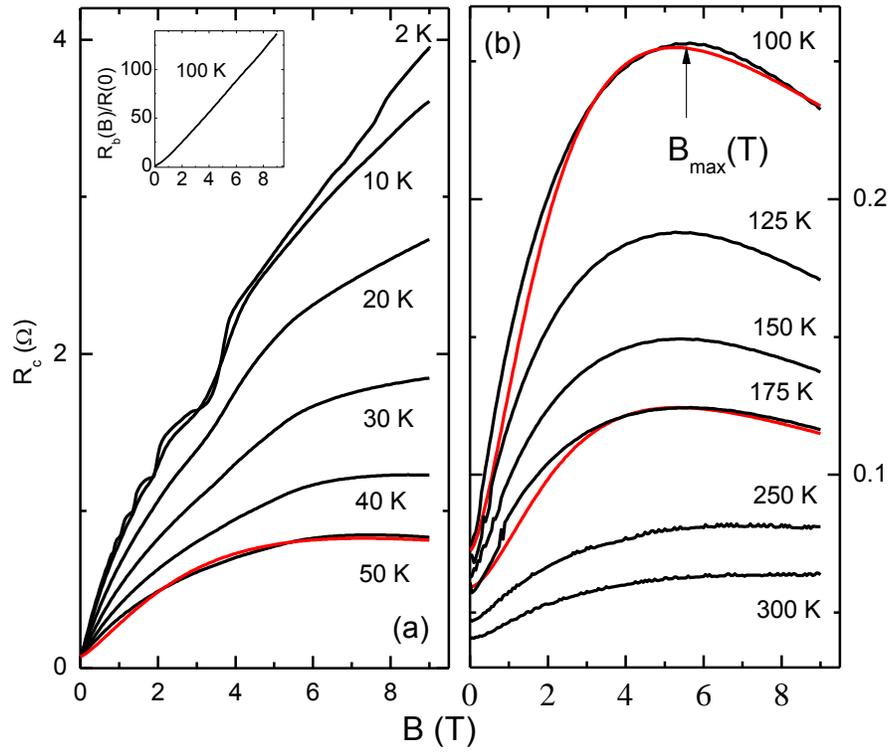

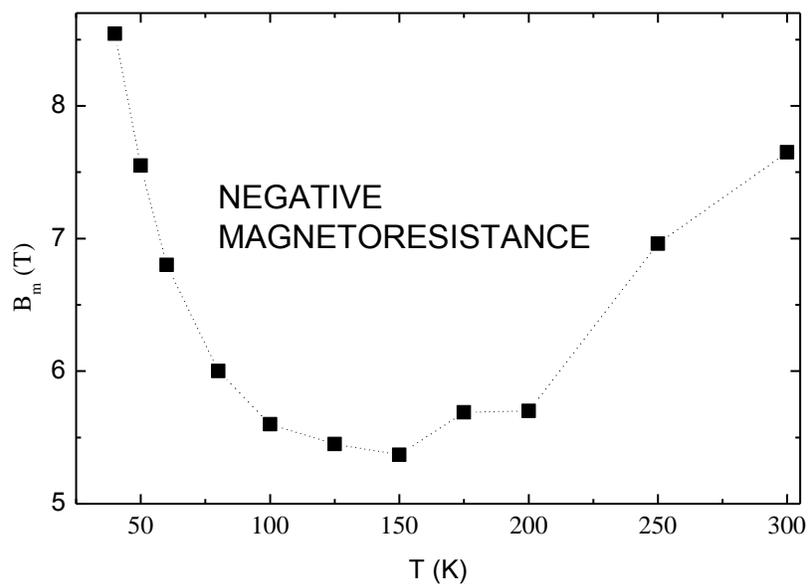

Fig. 2

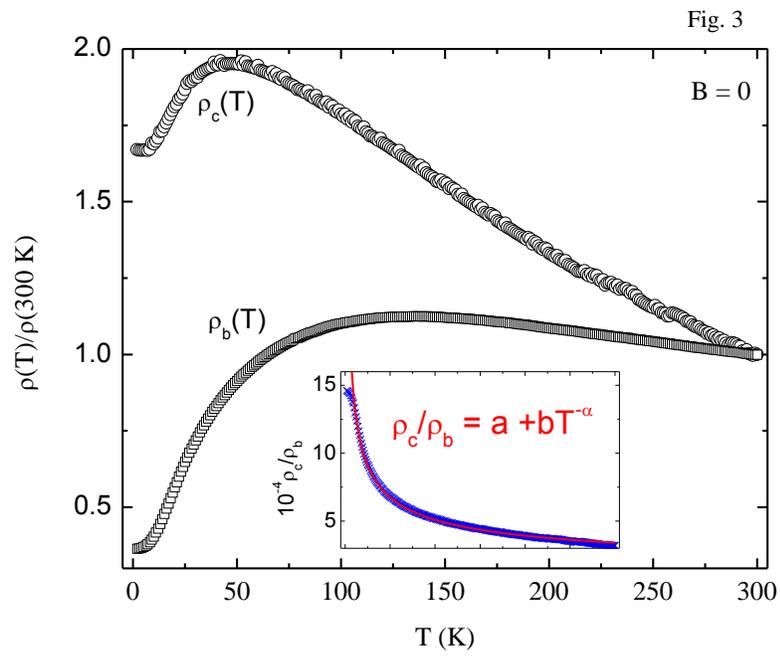

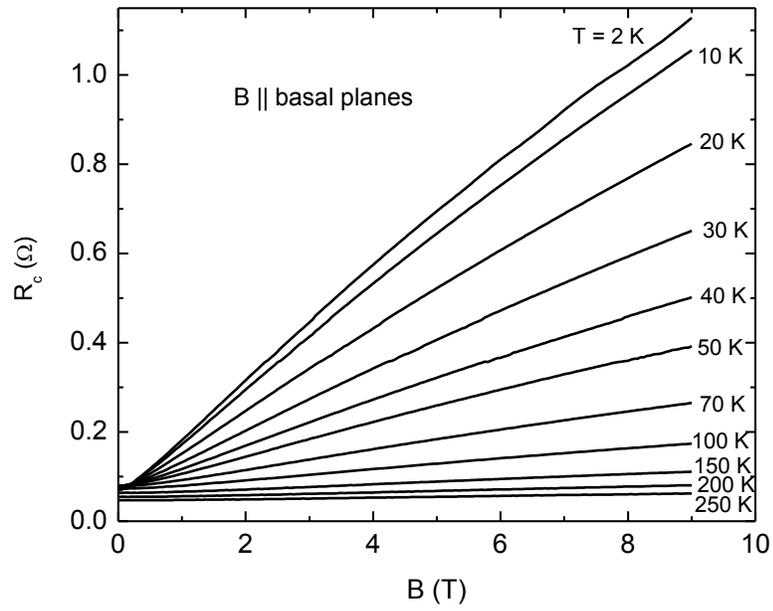

Fig. 4